\documentclass[%
 aip,
 amsmath,amssymb,
reprint,
twocolumn]{revtex4-2}

\usepackage{graphicx}
\usepackage{dcolumn}
\usepackage{bm}
\usepackage{upgreek}
\usepackage[utf8]{inputenc}
\usepackage{etoolbox}
\usepackage{xcolor}
\usepackage{multirow}
\usepackage{nicematrix}
\NiceMatrixOptions{cell-space-limits=1pt}

\usepackage{array}
\usepackage{booktabs}
\usepackage{siunitx}
\usepackage[margin=1in]{geometry}
\usepackage{caption}
\usepackage{pict2e}

\makeatletter
\def\@email#1#2{%
 \endgroup
 \patchcmd{\titleblock@produce}
  {\frontmatter@RRAPformat}
  {\frontmatter@RRAPformat{\produce@RRAP{*#1\href{mailto:#2}{#2}}}\frontmatter@RRAPformat}
  {}{}
}%
\makeatother
\begin{document}
\renewcommand{\arraystretch}{1.4} 
\preprint{AIP/123-QED}

\title{Robust quantum Hall resistance standard \\from uniform wafer-scale epitaxial graphene on SiC}

\author{François Cou\"edo}
\affiliation{ 
Laboratoire National de Métrologie et d’Essais (LNE), 29 Avenue Roger Hennequin, 78197 Trappes, France
}%
  \email{francois.couedo@lne.fr}
  
\author{Chiara Mastropasqua}
\affiliation{%
Université Côte d’Azur, CNRS, Centre de Recherche sur l'Hétéroépitaxie et ses applications (CRHEA), rue Bernard Grégory, Valbonne, France.
}%

\author{Aurélien Theret}
 \altaffiliation{Also at Laboratoire National de Métrologie et d’Essais (LNE). Now at CERFACS, Toulouse, France.}
\affiliation{%
Université Paris-Saclay, CNRS, Centre de Nanosciences et de Nanotechnologies (C2N), Palaiseau, France
}%

\author{Dominique Mailly}
\affiliation{%
Université Paris-Saclay, CNRS, Centre de Nanosciences et de Nanotechnologies (C2N), Palaiseau, France
}%

\author{Adrien Michon}
\affiliation{%
Université Côte d’Azur, CNRS, Centre de Recherche sur l'Hétéroépitaxie et ses applications (CRHEA), rue Bernard Grégory, Valbonne, France.
}%

\author{Mathieu Taupin}%
\affiliation{ 
Laboratoire National de Métrologie et d’Essais (LNE), 29 Avenue Roger Hennequin, 78197 Trappes, France
}%

\date{\today}

\begin{abstract}

We report high-precision resistance measurements on quantum Hall resistance devices fabricated from uniform epitaxial graphene grown by propane-hydrogen chemical vapor deposition on a two-inch silicon carbide substrate. Through molecular doping, we achieve a low carrier density regime ($n_\mathrm s < $ 1.5 \textperiodcentered 10$^{11}$ cm$^{-2}$) combined with high mobility ($\upmu \geq$ 6000 cm$^2$ V$^{-1}$ s$^{-1}$) at low temperature. Accurate quantization of the Hall resistance is demonstrated at magnetic flux densities as low as 3.5 T, temperatures up to 8 K, and measurement currents up to 325 $\upmu$A, with relative measurement uncertainties of a few parts per billion. A stability diagram mapping dissipation as a function of temperature and current provides insight into optimal doping conditions that maximize the breakdown current. All measurements were carried out in a pulse-tube-based cryomagnetic system, enabling simplified and continuous operation of the quantum Hall resistance standard without liquid helium consumption.

\end{abstract}

\maketitle


The broader dissemination of electrical units of the International System of Units (SI) requires the development of quantum electrical standards that can be implemented in compact and user-friendly cryogenic systems. Epitaxial graphene on silicon carbide (G/SiC) has emerged as the leading material platform for the next-generation quantum Hall resistance standard (QHRS), enabling exact quantization of the Hall resistance at filling factor $\upnu = 2$ under more relaxed experimental conditions than conventional GaAs-based heterostructures \cite{ribeiro2015quantum}. Despite this advantage, widespread adoption of graphene-based QHRS still faces technological challenges - most notably the scalable growth of highly homogeneous monolayer G/SiC and the precise control of doping - before their potential can be fully exploited in resistance metrology, but also in current \cite{kaneko2023perspectives,djordjevic2025primary}, impedance \cite{kalmbach2014towards,overney2024longitudinal} and even mass \cite{thomas2017determination} metrology.

\begin{figure*}[t]
\includegraphics[width=0.8\textwidth]{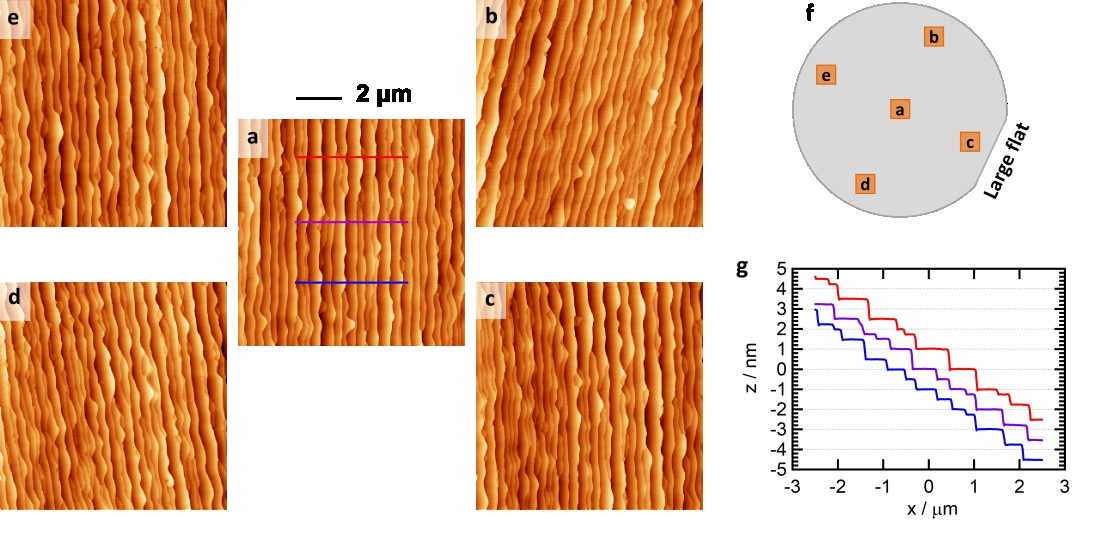}
   \caption{\textbf{Structural characterization}. AFM images in the center (a) and in the periphery (b-e) of a graphene grown on a two-inch SiC wafer, as shown in (f). AFM images are tilted by 25° with respect to SiC large flat, and with a $z$-scale of 1.5 nm. (g) Line profiles in the center.}
\label{fig:SampleCarac}
\end{figure*}

The scalable fabrication of millimeter-sized Hall bar devices for quantum Hall metrology requires large-area and highly homogeneous G/SiC. Both propane-hydrogen chemical vapor deposition (CVD) \cite{michon2013effects,liu2018chemical,ribeiro2015quantum,lafont2015quantum} and polymer-assisted sublimation growth (PASG)\cite{kruskopf2016comeback} have demonstrated the ability to produce high-quality G/SiC suitable for metrological application \cite{ribeiro2015quantum,lafont2015quantum, yin2022tailoring, yin2024quantum}. Nonetheless, achieving wafer-scale graphene with uniform electronic properties remains a key challenge.

To enable operation of the QHRS in compact cryocooler-based systems at moderate magnetic fields ($B \leq 5$ T) and temperature ($T \geq$ 4 K), precise control of the carrier density ($n_\mathrm{s}$) is essential. In pristine G/SiC, $n_\mathrm{s}$ is typically about $10^{13}$ cm$^{-2}$ due to charge transfer from the buffer layer \cite{kopylov2010charge}. Among the available approaches \cite{lara2011non, lartsev2014tuning, rigosi2019gateless}, molecular doping is currently one of the most promising technique for reducing $n_\mathrm{s}$ with reproducibility and homogeneity \cite{he2018uniform, yin2022tailoring,shetty2023long,yin2025graphene}. However, its long-term stability remains inferior to that of GaAs-based quantum Hall resistance (QHR) devices, which are operated for decades in National Metrology Institutes. To address this challenge, strategies to minimize temporal drift of doping are currently being developed, such as storage under controlled conditions \cite{park2022glass, shetty2023long} or device encapsulation \cite{hu2018towards, park2022glass}.

Here, we demonstrate metrology-grade QHR devices based on a graphene uniformly grown on a two-inch SiC substrate. We achieve reproducible carrier density control through molecular doping and evaluate device performance for more than two years in terms of doping homogeneity and stability, as well as accuracy of the Hall resistance quantization as a function of magnetic field, temperature, and measurement current. 

Cryogenic measurements were performed in a cryomagnetic system from Cryogenic Limited combining a pulse-tube cryocooler, a $^4$He variable temperature insert (VTI), a 14 T superconducting magnet and a top-loading cryoprobe designed for precision electrical measurements. This system works without liquid helium consumption, offering a cost-effective and simplified alternative to conventional liquid-helium-based cryostats. The characterization of this system for metrological electrical measurements is detailed in Ref.~\cite{taupin2025}. High-precision direct current (dc) measurements of the longitudinal resistance $R_\mathrm{xx}$ were carried out with an N31 nanovoltmeter from EM Electronics connected to two voltage terminals of the Hall bar, while a dc current $I$ is supplied from a low-noise homemade current source \cite{poirier2020resistance}. For dc precision measurements of the Hall resistance $R_\mathrm{H}$, we used a homemade resistance bridge based on a cryogenic current comparator (CCC) \cite{poirier2020resistance}. 

Graphene was grown by propane-hydrogen CVD \cite{mastropasqua2025self} on the Si-face of a semi-insulating on-axis 4H-SiC two-inch substrate from PAM-Xiamen using a home-made horizontal hot wall CVD reactor. A hydrogen-argon mixture (25 \% of hydrogen) at a pressure of 800 mbar was used as the carrier gas for the temperature ramp and for the 15 minutes growth plateau at 1550 °C. With hydrogen in the gas phase, no carbon excess can be obtained on the SiC surface at this temperature \cite{dagher2018comparative}. Hence, to grow graphene, a propane ﬂow (0.83 \%) was added during the 15 minutes growth plateau, after which cooling was carried out under argon only, still at a pressure of 800 mbar. These conditions lead to the growth of graphene on a buffer layer with a self-limited monolayer thickness \cite{mastropasqua2025self}.


Figures~\ref{fig:SampleCarac}(a-e) present atomic force microscopy (AFM) images taken at representative positions across the wafer, indicated in Fig.\ref{fig:SampleCarac}(f). These images reveal smooth terraces with typical widths of 400 nm and up to 800 nm. In Fig.\ref{fig:SampleCarac}(g), the $z$-profiles show step heights that are multiples of 0.25 nm, corresponding to the thickness of a Si-C bilayer, and with a maximum of 1 nm (equal to the $c$-axis lattice parameter of 4H-SiC). These highly uniform structural properties over the entire SiC wafer are therefore consistent with self-limitated graphene monolayer growth over large scale, as demonstrated in our previous study \cite{mastropasqua2025self}. Moreover, the AFM images also show no significant step bunching, as usually observed in graphene grown by silicon sublimation. The suppression of step bunching - characteristic of both CVD under hydrogen and PASG \cite{kruskopf2016comeback} - can be attributed to the presence of an external carbon source.


\begin{table*}[t]
\centering
\caption{\textbf{Electrical properties from Hall effect.} Carrier density ($n_\mathrm s$) and mobility ($\upmu$) for 5 samples made from a single wafer, before and after doping, both at room temperature (RT) and low temperature (LT : $T \leq $ 5 K).}
\label{table-MR_lowfield}
\begin{NiceTabular}{lccccc}[hvlines]
\toprule
   & \textbf{02d} & \textbf{02g} & \textbf{04d} & \textbf{03g} & \textbf{04g} \\
  \midrule
 \textbf{Before doping (RT)} \\
$n_\mathrm s$ (cm$^{-2}$) & 5.2 $\cdot$ 10$^{12}$ & 4.8 $\cdot$ 10$^{12}$ & 3.9 $\cdot$ 10$^{12}$ & \multicolumn{2}{c}{} \\
 $\upmu$ (cm$^2$ V$^{-1}$s$^{-1}$) & 1 429 & 1 554 & 1 545 & &\\ 
  \midrule
  
\textbf{After doping (RT)} \\
$n_\mathrm s$ (cm$^{-2}$) & 3.8 $\cdot$ 10$^{11}$  & 4.9 $\cdot$ 10$^{11}$ & 3.0 $\cdot$ 10$^{11}$ & \multicolumn{2}{c}{} \\
$\upmu$ (cm$^2$ V$^{-1}$s$^{-1}$) & 3 178 & 4 040 & 3 252 & & \\
\midrule

\textbf{After doping (LT)} \\
$n_\mathrm s$ (cm$^{-2}$) & 7.2 $\cdot$ 10$^{10}$ & 1.34 $\cdot$ 10$^{11}$ & 1.42 $\cdot$ 10$^{11}$ \textit{(holes)} & 1.05 $\cdot$ 10$^{11}$ & 6.6 $\cdot$ 10$^{10}$ \\
$\upmu$ (cm$^2$ V$^{-1}$s$^{-1}$) & 7 023 & 18 528 & 6 100 & 14 038 & 22 926 \\
\bottomrule
\end{NiceTabular}
\end{table*}

\begin{figure*}[t]
\includegraphics[width=0.75\textwidth]{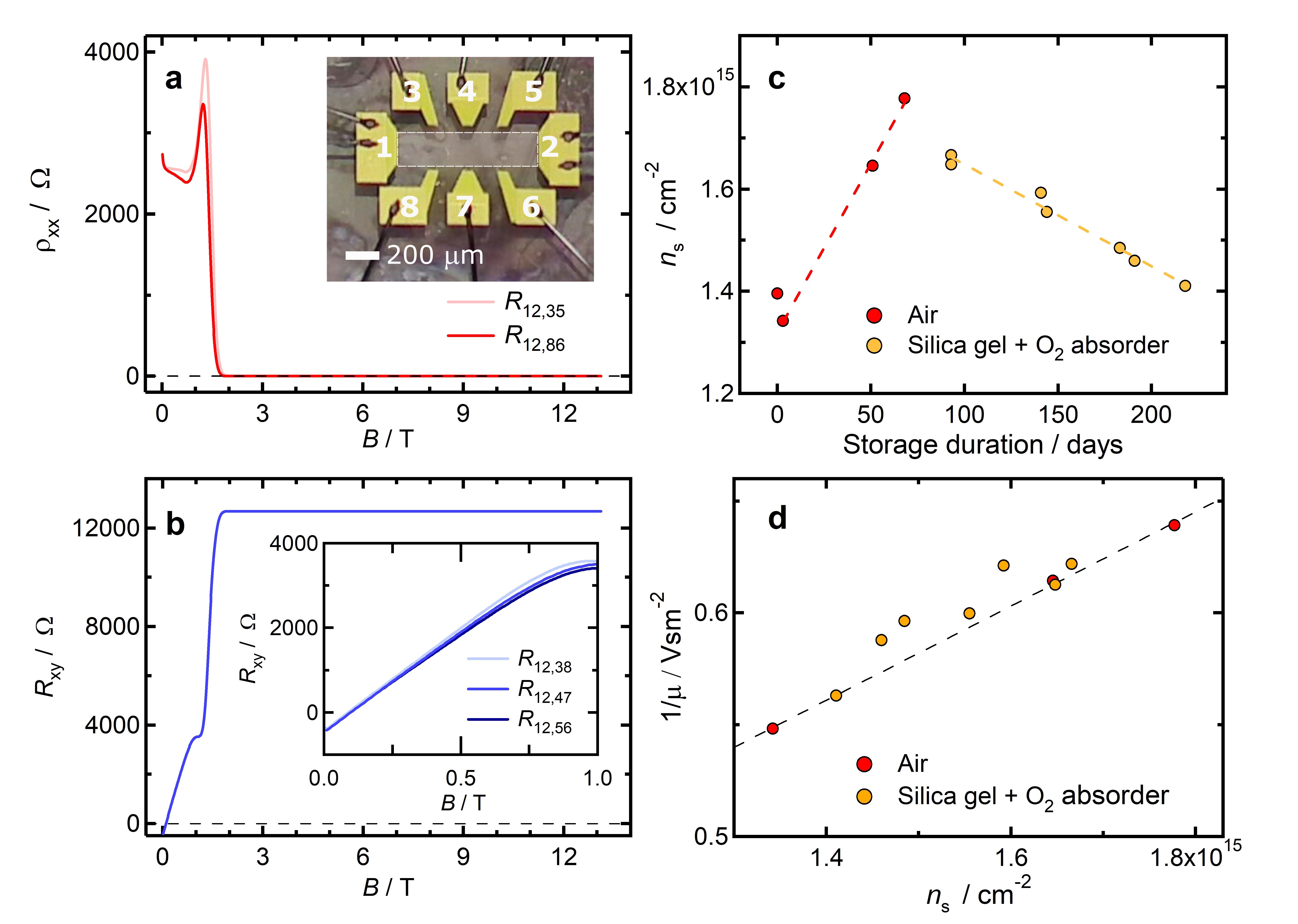}
\caption{\textbf{Magnetotransport and temporal stability after molecular doping (sample 2g)}. (a) Longitudinal resistivities and (b) Hall resistances as a function of the magnetic induction. The inset of (a) shows an optical image of the Hall bar. (c) Low-temperature carrier density $n_\mathrm s$ as a function of the storage duration of the sample at room temperature in the laboratory (relative humidity and temperature controlled), in air (red) and successively in a box containing silica gel and O$_\mathrm 2$ absorber (yellow). (d) Inverse of the corresponding mobility as a function of the carrier density.}\label{fig-MR}
\end{figure*}

The wafer is diced into 5 mm x 10 mm pieces, each of them defining a sample including up to 8 Hall bars. Prior to the device processing, each sample is annealed in a few 10$^{-6}$ mbar vacuum at 450 $^{\circ}$C for 2 hours to remove any organic residual impurities. All further processings use PMMA resist and electron beam lithography. The mesa is delimited using oxygen reactive etching. Ohmic contacts to the graphene layer are performed with a Ti(5 nm)/Pd(20 nm)/Au(100 nm) trilayer lift off using electron beam deposition system. Finally, electron beam lithography is performed to uncover the contact pads from PMMA for wire bounding on the chip carrier. Hall bars are 200 $\upmu$m wide with a length of 200 $\upmu$m between two longitudinal probes. The overall length between source and drain is 1 mm [Inset of Fig.~\ref{fig-MR}(a)]. 
   
Hall effect measurements are first performed at room temperature using a probe station, prior to any molecular doping. In total, 28 Hall bars from 4 different samples were electrically characterized. The average carrier density and mobility were found to be $\overline{n}= 4.7\cdot10^{12}$ cm$^{-2}$ and $\overline{\upmu}=$ 1404 cm$^2$ V$^{-1}$ s$^{-1}$, respectively. The dispersion for both quantities was less than 10$\%$, indicating excellent homogeneity of the electronic properties across the wafer, consistently with the structural uniformity observed by AFM.

To reduce the intrinsic $n-$type doping of G/SiC, we used 2,3,5,6-Tetrafluor7,7,8,8-tetracyanoquinodimethane (F4-TCNQ) molecules, which act as an effective $p-$type dopant for graphene \cite{he2018uniform}. 
Molecular doping was performed using a three-layer doping stack, where a single F4-TCNQ doping layer is separated from the graphene monolayer by one spacer layer and with one cap layer on top. The process is established according to the following steps. Dry powder of F4-TCNQ (25 mg) is first dissolved in 3 mL of anisole and 0.5 mL of this solution is then mixed with 1 mL of PMMA to realize the so-called "dopant blend". After the pre-deposition of a 100-nm-thick PMMA buffer layer on G/SiC, successively baked at 160 $^{\circ}$C during 5 min, the dopant blend is spin coated, resulting in a 150-nm-thick-layer. The sample is then baked at 160 $^{\circ}$C during 5 min. Finally, the sample is covered with a 350-nm-thick layer of PMMA for protection. No post-deposition annealing was performed, although such treatment could further provide an additional level of fine tuning the doping \cite{yin2022tailoring}. 

At room temperature, the F4-TCNQ deposition reduces $n_\mathrm s$ by approximately one order of magnitude [Table~\ref{table-MR_lowfield}]. This reduction becomes even more pronounced at low temperature ($T \leq $ 5 K), where $ \lvert n_\mathrm s \rvert <$ 1.5 \textperiodcentered 10$^{11}$ cm$^{-2}$. Importantly, each of the five samples characterized at low temperature underwent an independent dopant deposition, which ensures the reproductivity of our molecular doping method. High doping homogeneity over millimeter-scale distances ($ \Updelta n_\mathrm s \slash n_\mathrm s \approx $ 3 $\%$) is moreover demonstrated by comparing the low-field Hall resistances measured across the three different voltage probe pairs within a given Hall bar [inset of Fig.\ref{fig-MR}(b)]. Furthermore, the high carrier mobilities ($ \upmu >$ 6 000 cm$^2$ V$^{-1}$ s$^{-1}$) reported in Table~\ref{table-MR_lowfield} confirm the low level of disorder in the graphene layer. In the following sections, we focus on a specific sample (labeled 2g), which has been extensively characterized for more than two years.

The stability of molecular dopants is known to depend sensitively on device storage conditions \cite{park2022glass,shetty2023long}. Figure~\ref{fig-MR}(c) shows the temporal evolution of the electron carrier density over the first 50 days of storing the device at room temperature in the highly controlled  environment of the laboratory at LNE (the durations at low temperature are not counted here). Under these conditions, the carrier density increased at a relative rate of approximately of 0.5 $\%$ per day. To mitigate this drift, oxygen absorbers and silica gel desiccants were added to a storage box with controlled temperature ($T \approx$ 20 $^{\circ}$C) and relative humidity RH (30 $\% <$ RH $<$ 35 $\%$), following strategies proposed in Ref.~\cite{shetty2023long}. After this change, the evolution of the doping reversed direction, moving toward the charge neutrality point, at a reduced rate of 0.1 $\%$ per day. Further studies are currently underway to assess whether the doping will eventually stabilize over longer timescales. 

The evolution of the corresponding mobility is plotted in inverted scale in Fig.\ref{fig-MR}(d), as a function of the carrier density. As expected for G/SiC \cite{he2018uniform}, the mobility decreases as $n_\mathrm s $ increases. Notably, the data from both storage conditions follow a nearly identical linear trend, indicating that the relationship between mobility and doping is independent of the environmental storage conditions. From this linear behavior, analysed within the charged impurity scattering model \cite{chen2008charged}, we extracted a density of charged impurities $n_\mathrm{imp} = 1.43$\textperiodcentered$ 10^{11}$ cm$^{-2}$ and a short-range resistivity $\uprho_s =$1254 $\Upomega$. These values, together with the reversibility of the drift, indicate that the temporal evolution of doping does not arise from variations in the underlying disorder - thus confirming the stability of intrinsic electronic properties over time - but rather suggest that it results from adsorption processes.


We now consider the properties in the quantum Hall regime. Figure~\ref{fig-MR} shows the magnetic flux density dependence of (a) the longitudinal resistivity $\rho_\mathrm{xx}$ ($= R_\mathrm{xx}\frac{W}{L}$, where $W$ and $L$ are the width and length between the probes, respectively) and (b) Hall resistance $R_\mathrm{H}$, measured at $T = 1.3$ K using low-frequency lock-in techniques. A clear quantum Hall plateau at $\upnu = 2$ emerges from $B\approx $ 2 T, where $\uprho_\mathrm{xx}$ vanishes and $R_\mathrm{H}$ approaches the quantized value $R_\mathrm{K}/2 = h/(2e^2)$, where $R_\mathrm{K}$ denotes the von Klitzing constant, defined in terms of the Planck constant $h$ and the elementary charge $e$. Following the technical guidelines for reliable dc QHR measurements \cite{delahaye2003revised}, we first verified the quality of the electrical contacts through three-terminal resistance measurements on this QHR plateau. All contact resistances were found to be sufficiently low, with measured values below 1.5 $\Upomega$; a value that includes the series resistance of the measurement leads, $R_\mathrm{leads} \approx$ 1.2 $\Upomega$.


\begin{figure*}[t]
\includegraphics[width=0.95\textwidth]{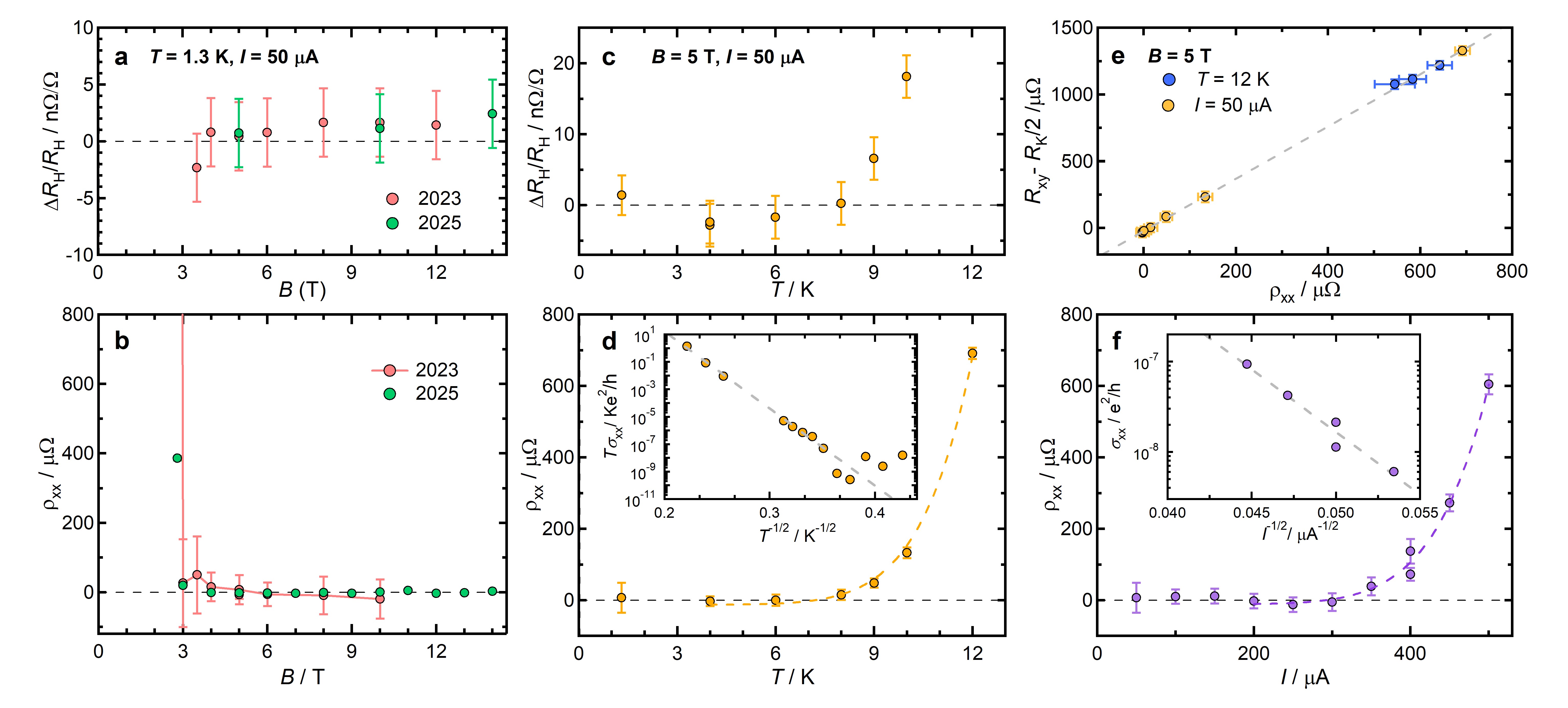}
\caption{\textbf{Precision resistance measurements (sample 2g).} Longitudinal resistivity $\uprho_\mathrm{xx}$ and Hall resistance $R_\mathrm{H}$ deviation from $R_\mathrm K /2$, expressed as the relative deviation $\Updelta R_\mathrm H / R_\mathrm H = \frac{{R_\mathrm{H}-R_\mathrm K /2}}{R_\mathrm K /2}$ as a function of magnetic induction (a) and (b), respectively, and as a function of temperature in (c) and (d), respectively. (e) Hall resistance deviation as a function of dissipation ($\uprho_\mathrm{xx}$). $\uprho_\mathrm{xx}$ as a function of measurement current in (f). The inset of (d) shows the product $T \sigma_{xx}$ as a function of $T^{-1/2}$. The inset of panel (f) shows the longitudinal conductivity $\upsigma_{xx}$ as a function of $I^{-1/2}$. The dashed lines in these insets represent fitting according to the VRH models in the dissipative regimes.}\label{fig:fig-quantif}
\end{figure*}

The accuracy of the Hall resistance quantization was tested through indirect comparison with a well-characterized GaAs-based QHRS by using a stable 100 $\Upomega$ transfer resistance standard. Figure~\ref{fig:fig-quantif}(a) shows the magnetic flux density range over which $R_\mathrm{H}$ remains accurately quantized, i.e. with a deviation from $R_\mathrm K/2$ smaller than our combined measurement uncertainty ($\Updelta R_\mathrm H / R_\mathrm H < $ 3\textperiodcentered 10$^{-9}$, $k=$1), which is dominated by the instability of the transfer standard. The corresponding longitudinal resistivity $\uprho_\mathrm{xx}$ is presented in Fig.\ref{fig:fig-quantif}(b). The quantization window spans from 3.5 T to (at least) 14 T, under conditions $T = 1.3$ K and $I =$ 50 $\upmu$A. Note that this perfect quantization was maintained during the two years of monitoring of the device, further attesting to the long-term stability of the technology. In the following, we set the magnetic flux density at a moderate value ($B = $ 5 T) and investigate the onset of dissipation, as a function of both temperature and current, to evaluate the robustness of the quantum Hall state.

Figures~\ref{fig:fig-quantif}(c) and (d) show the evolution of $R_\mathrm{H}$ and $\uprho_\mathrm{xx}$ as a function of temperature, measured at $I = 50$ $\mu$A, respectively. The accurate quantization of $R_\mathrm{H}$ is preserved up to $T =$ 8 K, consistent with the very low dissipation observed in this regime ($\uprho_\mathrm{xx} \leq$ (15 $\pm$ 14) $\upmu \Upomega$). This performance surpasses the former high-temperature limit for Hall resistance quantization ($T=$ 5 K, under same current and field conditions) \cite{ribeiro2015quantum}. This exceptional robustness at elevated temperature is further supported by an analysis based on the variable range hopping theory (VRH) with Coulomb interactions \cite{polyakov1993variable}. As shown in the inset of Fig.\ref{fig:fig-quantif}(d), the experimental data are well described by this model, which predicts a temperature dependence of the longitudinal conductivity of the form $\upsigma_\mathrm{xx} \propto \frac{1}{T}\mathrm{exp}[-(T_\mathrm 0/T)^{1/2}]$. In the low-current limit ($I = $100 nA), we extract a characteristic temperature $T_\mathrm 0 = $ (1573 $\pm$ 46) K (the uncertainty here only reflects spatial inhomogeneities). From this value, a localization length $ \upxi = $ (9.1 $\pm$ 0.3) nm is obtained from the formula $ \upxi = Ce^2/4 \uppi k_\mathrm B T_\mathrm 0 \upepsilon_r \upepsilon_\mathrm 0 $ \cite{furlan1998electronic}, where $C = $ 6.2, $ k_\mathrm B$ is the Boltzmann constant, $ \upepsilon_\mathrm 0 $ is the permittivity of free space and $ \upepsilon_r $ is the mean relative permittivity of the G/SiC covered by the PMMA and dopant layers. It is noteworthy that the strong localization regime ($ \upxi \sim l_\mathrm B$, where $l_\mathrm B =\sqrt{\hbar/eB} $ is the magnetic length, the fundamental length scale in the presence of a magnetic field \cite{goerbig2011electronic}) is achieved here at a moderate magnetic field \cite{lafont2015quantum}, which further supports the robustness of the quantum Hall state.

\begin{figure*}[t]
\includegraphics[width=0.7\textwidth]{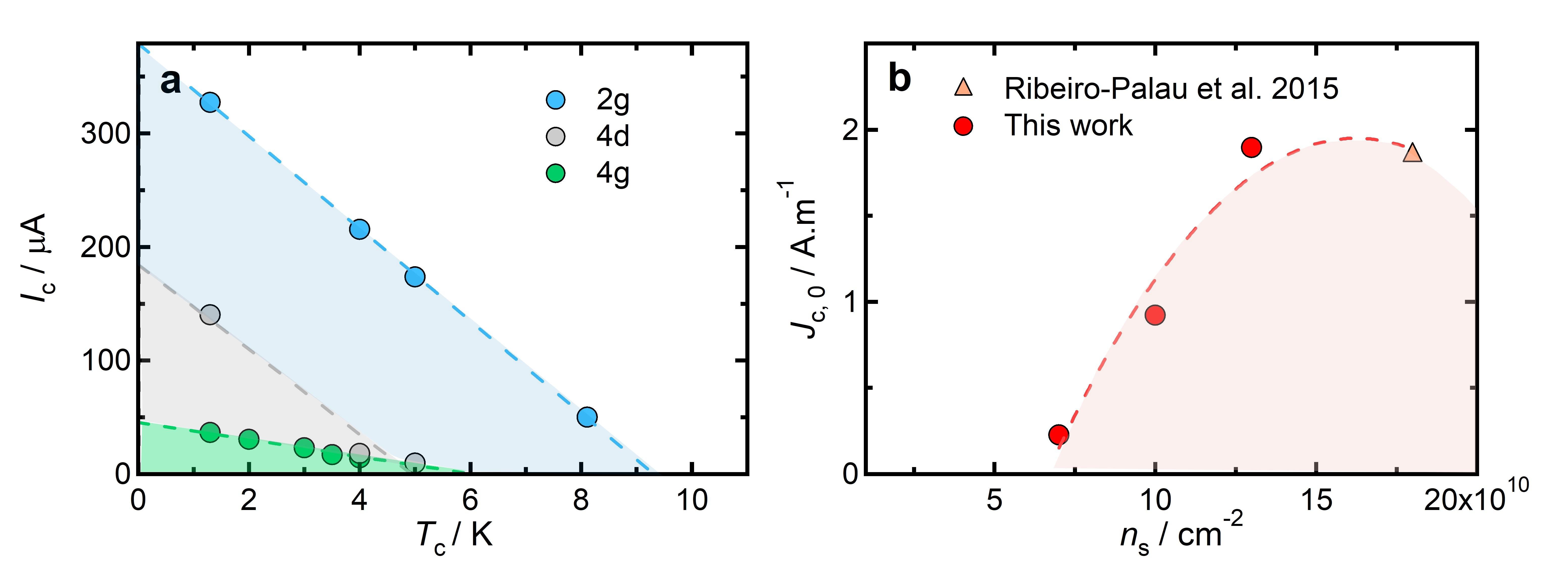}
\caption{\textbf{Stability diagrams at $B =$ 5 T.} (a) Evolution of the breakdown current $I_\mathrm c$ as a function of temperature. Coloured regions emphasize the low dissipation regimes ($\uprho_\mathrm{xx} \leq $ 20 $\mu \Upomega$). (b) Evolution of the breakdown current density, extrapolated at zero temperature, $J_\mathrm {c,0}$, as a function of the carrier density, for three samples from this work (2g, 4g, 4d, red dots) as well as data from Ref.~\cite{ribeiro2015quantum}. The dashed line is a guide for the eye.}\label{fig:fig-highCurrent}
\end{figure*}

 The current dependence of the longitudinal resistivity at $T = 1.3$ K is shown in Fig.\ref{fig:fig-quantif}(f). A pronounce increase in $\uprho_\mathrm{xx}$ is observed for measurement currents larger than 325 $\mu$A, marking the onset of the breakdown regime. This value corresponds to a high breakdown current density $J_\mathrm c = I_\mathrm c/W$ of 1.63 A\textperiodcentered m$^{-1}$. We will come back later on the discussion of this value, and notably its relation with doping. Note that the current-induced dissipation is well described by an exponential growth characteristic of the VRH theory under high electric field, which predicts a relation of the form $\upsigma_\mathrm{xx} \propto \mathrm{exp}[-(I_0/I)^{1/2}]$, as shown on the inset of Fig.\ref{fig:fig-quantif}(f).

We now analyze how the Hall resistance deviates from its nominal value in the high-current regime — a measurement range not accessible with our resistance bridge — by determining the coupling factor $s$ that links this deviation to the dissipation, using the empirical relation: $\Updelta R_\mathrm H = R_\mathrm{H}- R_\mathrm K /2 = s \uprho_\mathrm{xx}$ \cite{cage1984temperature, chae2022investigation}. Figure~\ref{fig:fig-quantif}(e) displays this parametric relation for two separate data sets: measurements at (i) a fixed current of 50 $\upmu$A across temperatures from 4 to 10 K, and (ii) a fixed temperature 12 K and currents ranging from 20 to 50 $\upmu$A. From a single linear fitting (dashed line), we extract a coupling factor $s = $ 1.96 that applies both when the dissipation is tuned by increasing current or temperature. Previous works have reported $|s|$-values ranging from 0.005 \cite{janssen2012precision} up to 0.67 \cite{lafont2015quantum}, with significant dispersion in between \cite{janssen2012precision,kruskopf2021graphene, chae2022investigation}. The high value we determined here questions the maximum possible value of $s$ that can be achieved in G/SiC, as well as the influencing factors at play.

Next, we determine the dissipation threshold above which the Hall resistance is no longer accurately quantized by combining the determined value of $s$ and the combined measurement uncertainty of our CCC-based resistance bridge. Based on this criterion, we identify a critical value of the longitudinal resistivity $\uprho_\mathrm{xx} \geq $ 20 $\upmu \Upomega$ above which the deviation in $R_\mathrm{H}$ becomes experimentally significant. Using this dissipation threshold, we extend the current-dependence analysis presented earlier at $T =$ 1.3 K [Fig.\ref{fig:fig-quantif}(f)], and thus extract the breakdown current $I_\mathrm c$ at different temperatures. Figure~\ref{fig:fig-highCurrent}(a) shows a linear decrease of $I_\mathrm c$ when the temperature is increased. From this linear evolution, we extract a breakdown current of $I_\mathrm c =$ 208 $\upmu$A at $T =$ 4.2 K, resulting in the combination ($B$, $T$, $J_\mathrm c$) = (5 T, 4.2 K, 1 A\textperiodcentered m$^{-1}$) close to the record (4.5 T, 4.2 K, 0.6 A\textperiodcentered m$^{-1}$) recently reported in Ref.\cite{yin2025graphene}.

 This analysis was extended to other samples from the same wafer. As shown in Fig.\ref{fig:fig-highCurrent}(a), the slopes describing the reduction of $I_\mathrm c$ with temperature are similar for samples 4d and 2g, but significantly smaller for sample 4g. Given the low carrier density of the latter ($ n_\mathrm s =$ 7\textperiodcentered 10$^{10}$ cm$^{-2}$), this likely marks the crossover to a regime where charge puddles dominate the onset of current-induced dissipation \cite{yang2016puddle}. Figure~\ref{fig:fig-highCurrent}(b) shows the breakdown current density extrapolated to zero temperature, $J_\mathrm {c,0}$, as a function of $n_\mathrm s$, including data from previous work on a higher density graphene grown in similar conditions \cite{ribeiro2015quantum}, for comparison. After a roughly linear increase of $J_\mathrm {c,0}$ with $n_\mathrm s$, a maximum appears near $n_\mathrm s =$ 1.5 \textperiodcentered 10$^{11}$ cm$^{-2}$, close to the doping value ($n_\mathrm s =$ 1.3 \textperiodcentered 10$^{11}$ cm$^{-2}$) reported by NPL that maximizes the breakdown current at $B = 5$ T \cite{janssen2015operation}. While a full understanding of the universal mechanisms limiting the breakdown current in G/SiC, at filling factor $\upnu = 2$, remains complex and requires further study, these results already provide practical guidance for optimizing doping to achieve high-current operation in QHR devices. 
 
  In conclusion, we demonstrate metrology-grade QHR devices fabricated from a large-scale, uniform graphene monolayer grown by propane-hydrogen CVD on a two-inch SiC wafer. We show accurate Hall resistance measurements in the limits of low magnetic flux density ($B$, $T$, $J_\mathrm c$) $=$ (4 T, 1.3 K, 0.25 A\textperiodcentered m$^{-1}$), high measurement current (5 T, 1.3 K, 1.63 A\textperiodcentered m$^{-1}$) and high temperature (5 T, 8 K, 0.25 A\textperiodcentered m$^{-1}$). Achieving such performance in a simplified cryogenic environment highlights the robustness of the technology and its potential for practical quantum metrology applications at higher currents and temperatures to support wider dissemination of the SI electrical units.

This work has been partially supported by the French National Research Agency (ANR) through the VanaSiC project (ANR-22-CE24-0022-01), the European Union’s Horizon Europe research and innovation programme through the Qu-Test project (HORIZON-CL4-2022-QUANTUM-05-SGA, under Grant Agreement No 101113901) and the Region Sud - Provence-Alpes - Côte d’Azur through the PlaGGe project. 

FC and MT thank F. Piquemal and W. Poirier for fruitful discussions and proof-reading.

\section*{AUTHOR DECLARATIONS}
\subsection*{Conflict of interest}{The authors have no conflicts to disclose.}
\subsection*{Author Contributions}{
François Cou\"edo and Mathieu Taupin contributed equally to this work.

\textbf{François Cou\"edo :} Conceptualization (equal); Investigation
(equal); Formal analysis (lead); Supervision (equal); Writing – original draft (lead); Writing – review \& editing (equal). \textbf{Chiara Mastropasqua :} Investigation (equal); Formal analysis (supporting). \textbf{Aurélien Theret :} Methodology (supporting); Investigation (supporting). \textbf{Dominique Mailly :} Methodology (lead); Investigation (equal); Supervision (equal); Writing – review \& editing (equal). \textbf{Adrien Michon :} Methodology (lead); Investigation (equal); Formal analysis (supporting); Supervision (equal); Writing – review \& editing (equal). \textbf{Mathieu Taupin:} Conceptualization (equal); Investigation (equal); Formal analysis (supporting); Supervision (equal) ; Writing – original draft (supporting); Writing – review \& editing (equal).}

\section*{Data Availability Statement}{The data that support the findings of this study are available from the corresponding authors upon reasonable request.}


\end{document}